\documentclass[12pt]{iopart}

\usepackage{iopams}  
\usepackage{graphicx}
\usepackage{dcolumn}
\usepackage{bm}

\begin{document}

\title{Gate-controlled conductance of superconducting NbN nanowires: coherent quantum phase-slip or Coulomb blockade?}

\author{M. S. Anwar$^{1,2,*}$ \& J. C. Fenton$^{1}$}

\address{$^1$ London Centre for Nanotechnology, University College London, 17 19 Gordon Street, London WC1H 0AH, U.K. \\ $^2$ Department of Mateirals Science and Metallurgy, University of Cambridge, CB30FS, Cambridg. U.K.}
\ead{msa60@cam.ac.uk}

\vspace{10pt}
\begin{indented}
\item[]April 2021
\end{indented}

\begin{abstract}
Coherent quantum phase slips are expected to lead to a blockade of dc conduction in sufficiently narrow superconducting nanowires below a certain critical voltage. We present measurements of NbN nanowires in which not only is a critical voltage observed, but also in which this critical voltage may be tuned using a side-gate electrode. The critical voltage varies periodically as the applied gate voltage is varied. While the observations are qualitatively as expected for quantum interference between coherent quantum phase slip elements, the period of the tuning is orders of magnitude larger than expected on the basis of simple capacitance considerations. Furthermore, two significant abrupt changes in the period of the variations during measurements of one nanowire are observed, an observation which constrains detailed explanations for the behaviour. The plausibility of an explanation assuming that the behaviour arises from granular Josephson junctions in the nanowire is also considered.
\end{abstract}


\section{Introduction}

\begin{figure}
\begin{center}
		\includegraphics[width=8cm]{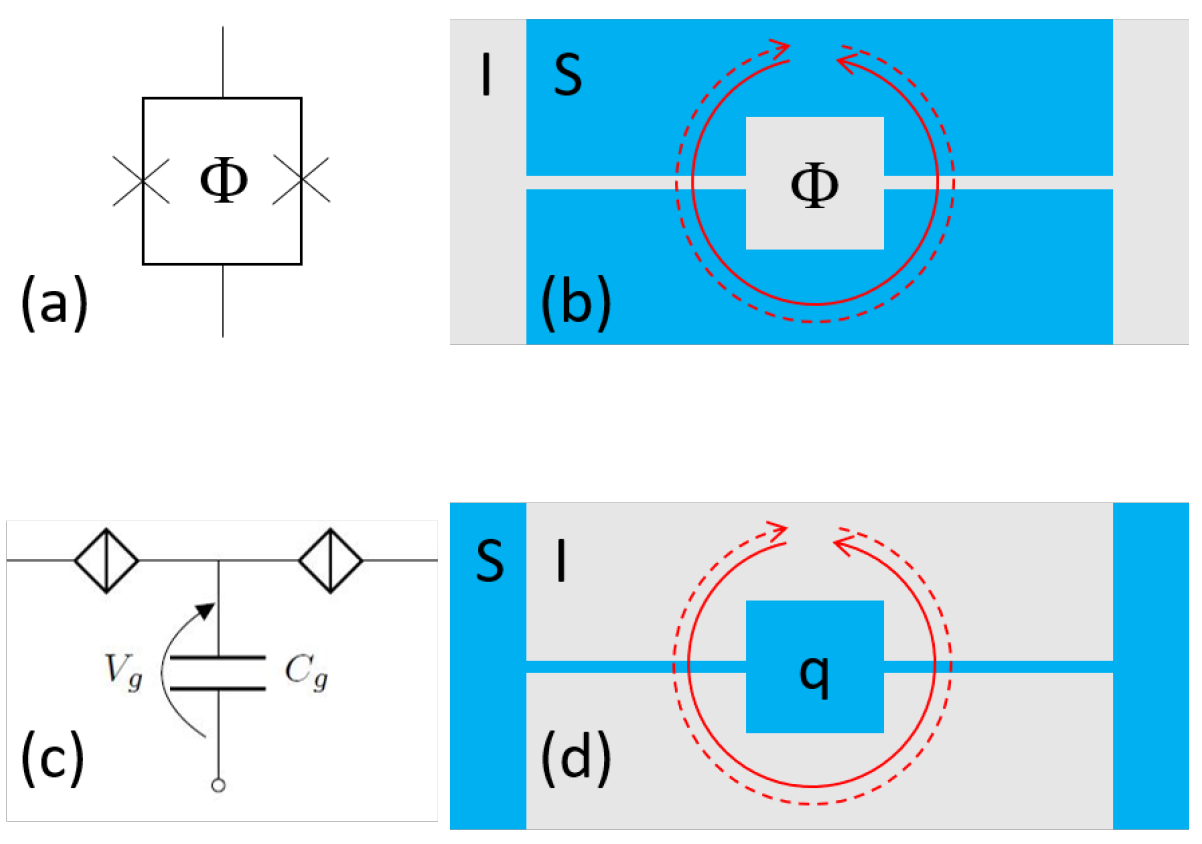}
		\caption{(a),(b): A dc-SQUID, formed of two Josephson junctions in parallel in a loop, with a flux applied to the loop by inductive coupling to the external environment. Interference
occurs between paths encircling the flux in opposite senses. (c),(d): A two-element CQPS device, with two CQPS elements in series, and a charge induced on the intervening island by capacitive coupling to the voltage on an external gate. Interference occurs between paths encircling the charge in opposite senses. (a),(c): Circuit representation. (b),(d): Schematic geometric layout with superconductor in blue and insulator/free-space light-coloured (after Ref.~6).}
		\label{Fig1}
	\end{center}
\end{figure}

Interference is one of the characteristic properties of wave like behaviour. It is seen in superconducting systems as the periodic magnetic field tuning of the maximum supercurrent through a dc superconducting quantum interference device (SQUID) [Fig.~\ref{Fig1}(a)] and is also seen in the variation of the maximum supercurrent of a wide Josephson junction as a function of in-plane magnetic field as a results of vector potential imposing a gradient along the junction in the superconducting phase-difference across the Josephson junction~\cite{Tinkham1996}; these phenomena are analogous respectively to the two-slit and Fraunhofer diffraction patterns familiar from optics.

A novel quantum superconducting system in which to consider interference behaviour is superconducting nanowires with large normal-state resistance and cross-sectional dimensions $\approx \xi_{\rm S}$, where $\xi_{\rm S}$ is the superconducting coherence length~\cite{Fenton2016}. Such nanowires undergo coherent quantum phase-slips (CQPS), involving coherent tunnelling of magnetic flux between the two sides of the nanowire, a process which is the charge-flux dual of Josephson tunnelling\cite{Mooij2006}. Particular interest and research efforts have been focussed on the potential of these nanowires to form the heart of a quantum current standard~\cite{Mooij2006,Webster2013}, single photon detectors \cite{Polakovic2020,Lyatti2020,Steinhauer2021} and it has also been suggested that CQPS elements are potentially useful in quantum computing~\cite{Friedman2002,Li2019}. It is furthermore of fundamental interest to investigate basic quantum phenomena in this quantum system $-$ for example, interference phenomena.

\begin{figure}
\begin{center}
		\includegraphics[width=8cm]{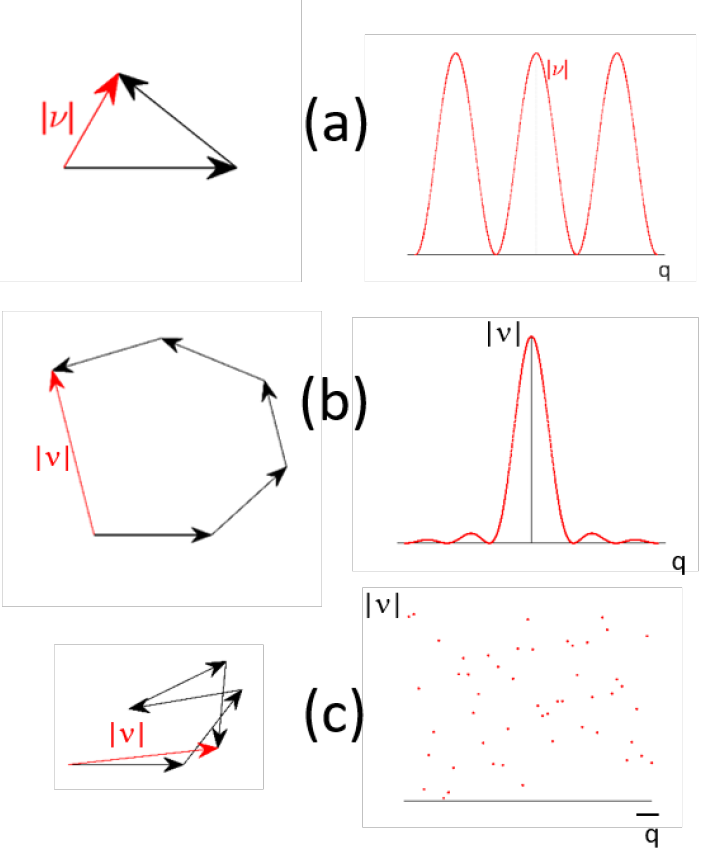}
		\caption{Sketches depicting summation of contributions from individual CQPS elements (black) to overall CQPS amplitude (red). (a) Two-element system, with identical elements. (b) Many-element system, with the same charge induced on all islands. (c) Many-element system, with random, uncorrelated charges induced on the islands (Left:) phasor diagram for one value of the tuning parameter $q$. (Right:) Variation of overall CQPS amplitude with $q$.}
		\label{Fig2}
	\end{center}
\end{figure}

The analogue of two parallel Josephson junctions in a dc SQUID inductively coupled to an external magnetic field is two CQPS elements in series separated by an island capacitively couple to the charge on a gate, which induces a charge $q$ on the island $-$ see Fig.~\ref{Fig1} (b). Interference between fluxon tunnelling paths that encircle the induced charge on the island should lead to a relative phase angle $2\pi q/2e$ between the two contributions to the QPS amplitude, giving a net amplitude and critical voltage ($V_{\rm c}$) which are 2$e$-periodic in the induced charge, as sketched in Fig.~\ref{Fig2}(a). For a larger number, $N$, of CQPS elements and islands in series and assuming the charge induced on the islands is correlated, similar periodic oscillations in net CQPS amplitude are expected, with maxima of decreasing magnitude as the induced charge increases, reaching a Fraunhofer dependence [Fig.~\ref{Fig2}(b)] in the limit of sufficiently large $N$. If, on the other hand, charges on successive islands were uncorrelated, the net CQPS amplitude would~\cite{Matveev2002} scale as $\sqrt{N}$ [Fig.~\ref{Fig2}(c)].

For wide study and application of CQPS phenomena, it is necessary to address two practical issues: maximising the CQPS amplitude of a CQPS element and the in influence of inhomogeneity. As an initial remark we note that, despite the practical challenges, there have been two previous experimental reports of CQPS phenomena in the geometry of two CQPS elements in series. Hongisto and Zorin~\cite{Hongisto2012} have reported experiments on NbSi nanowires interrupted by a gate showing just such oscillations in $V_{\rm c}$ as a function of gate voltage ($V_{\rm g}$), and simple considerations relate the period in $V_{\rm g}$ to the 2$e$-periodicity of the island charge via the capacitance between the gate and the island. De Graaf {\it et al}~\cite{Graaf2018} have also recently reported spectroscopy on a short-nanowire sample in this two element geometry, demonstrating the expected avoiding crossing between energy levels as a result of the gate tunable CQPS amplitude.

The magnitude of CQPS-related phenomena scales with the CQPS amplitude and therefore it is important to maximise the CQPS amplitude of each CQPS element. The primary way to achieve this for nanowires is through controlling the normal-state resistance of the used material and the cross-sectional dimensions, upon each of which the CQPS amplitude depends exponentially, and each of which requires careful tuning. The CQPS amplitude scales only linearly with the length of the nanowire, but increasing the length by orders of magnitude is straightforward~\cite{quote1}.

It is often argued~\cite{Vanevic2012} that inhomogeneity in real nanowires will mean that CQPS will be dominated by short sections of the actual nanowire, implying that a simple proportionality of the CQPS amplitude on the length in a homogeneous nanowire is not relevant to real nanowires. Instead, a long nanowire could behave as a series of CQPS elements with interposing islands, with a net CQPS amplitude determined by interference according to the charges on these islands. Reports of CQPS in nanowires with length $>>\xi_{\rm S}$ have been limited and further data in this area is called for.
 
Inhomogeneity in the materials also gives rise to an issue that is fairly fundamental to the development of applications based on CQPS in superconducting nanowires. This is the apparent close correspondence in the phenomenology of CQPS to the phenomenology arising from Coulomb blockade physics. We discuss this question in the next section.

\section{CQPS phenomena vs Coulomb blockade (CB)}
CQPS arises in a homogeneous nanowire containing no Josephson junctions but experiencing superconducting fluctuations related to its small cross-section~\cite{Arutyunov2008}. However, amorphous or nanocrystalline materials are generally employed to achieve a superconductor with a high normal-state resistance; since grain boundary Josephson junctions could form in these, it is a $priori$ plausible that a superconducting nanowire in these materials could contain one or many grain boundary Josephson junctions and these could be conjured in either a one-dimensional chain, or a more complex two-dimensional network. In a small Josephson junction, the charging energy scale $E_{\rm C}$ can exceed the Josephson energy $E_{\rm J}$, leading to the superconducting charge being well defined and the superconducting phase difference being undefined~\cite{Averin1985}. At sufficiently low junction bias, this can lead to a CB preventing charge transport~\cite{Haviland1991}. The CB of transport in Josephson junctions can lead to qualitatively similar behaviour to that expected due to CQPS. One dimensional chains of Josephson-junctions have also been studied theoretically and also measured experimentally~\cite{Haviland1991,Hermon1996,Zimmer2013,Vogt2015,Voss2021}. 

In nanowires, a current blockade was observed in titanium nanowires by Schollmann {\it et al}.~\cite{Schollmann2000}, with $V_{\rm c}$ tuned by a $V_{\rm g}$. These observations were attributed to CB. Similar observations, interpreted likewise as arising from CB, have also been reported in indium oxide nanowires by Chandrasekhar and Webb~\cite{Chandrasekhar1994} and in non-superconducting chromium nanowires by Krupenin {\it et al}.~\cite{Krupenin2001}. A recent preprint relating the properties of films and nanowires to the superconductor-insulator transition similarly interprets results on nanowires in a framework based on granular Josephson junctions~\cite{Schneider2018}.

\begin{figure}
\begin{center}
		\includegraphics[width=12cm]{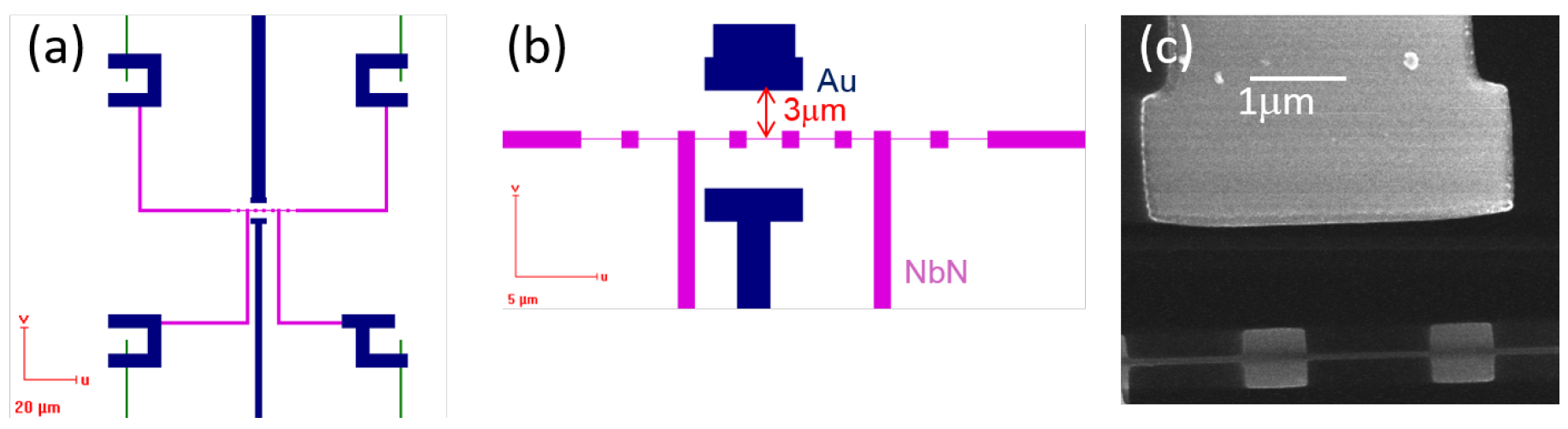}
		\caption{Nanowire NbN100/6. (a), (b) Design used to create the device. Pink elements are fabricated in niobium nitride, while blue elements are gold. (c) Helium-FIB micrograph of the device. A lift-off error (not shown) means that the lower gate is shorted to the left lower (voltage) electrode. $V_{\rm g}$s were applied only to the unaffected upper gate.}
		\label{Fig3}
	\end{center}
\end{figure}

The qualitative similarity between behaviour in the physics of CQPS on the one hand and the CB on the other is present in several properties of these devices: a blockade of current at low bias arises both from CQPS in a superconducting nanowire and from the CB in a tunnel junction. Current steps may be expected in a tunnel junction device under microwave irradiation as well as in a superconducting nanowire device under microwave irradiation
and the single-electron transistor tunes with an applied $V_{\rm g}$ in a similar way to the two-element CQPS device. Although hardly immediately apparent from the literature relating to QPS effects in superconducting nanowires, there is an ongoing understated discussion within the community about whether some (or all) previously reported CQPS effects actually arise from CB effects in granular Josephson junctions. It has further been argued by some that
the similarities in phenomenology are present because the two pictures are simply different ways of describing precisely the same coherent-flux-tunnelling phenomenon; while this issue remains unsettled, the similarities at least mean that there are considerable challenges for experimentalists
seeking categorically to distinguish between the two possible underlying causes of observed behaviour. To date, justifications for interpretations of behaviour in terms of nanowire CQPS have largely relied on the better correspondence of extracted parameters to those expected in the CQPS scenario rather than the (granular) Josephson junction scenario.

In this paper, we study the properties of two NbN nanowires with length 8$-$10 $\mu$m which show a low temperature current blockade below a $V_{\rm c}$. We present transport measurements as the voltage applied to a side gate is varied; we demonstrate that the application of the side-$V_{\rm g}$ leads to periodic oscillations of the $V_{\rm c}$. The details of this tuning provide clues to the source of the behaviour.

\section{Experimental}
Superconducting NbN nanowires were fabricated via subtractive processing from 20-nm-thick NbN films deposited by dc magnetron sputtering. Nanowires were patterned in hydrogen silsesquioxane (HSQ) negative resist by e-beam lithography (EBL) patterning and reactive ion-etching was used to transfer the pattern into the films. The ultimate width obtainable for resulting nanowires is $\approx$~20 nm. Wider (200~nm wide) sections of nanowire were included in series with the narrow nanowire to provide additional kinetic inductance. Nanowire NbN81/1, shown in Fig.~\ref{Fig7}, is a 10.5~$\mu$m-long 50-nm-wide \cite{quote2} nanowire and has 200~$\mu$m of wider inductive nanowire in series. Nanowire NbN100/6, shown in Fig.~\ref{Fig3}, consists of 8~$\mu$m of NbN nanowire with width $\approx$~40 nm in 1.3-2.2-$\mu$m long sections separated by 1-$\mu$m-square islands. These islands were included both in order to act as well-defined islands onto which charge may be induced and also as mechanical anchors to mitigate issues caused by weak
adhesion of the nanowire during processing~\cite{Constantino2018}. Thin film chromium oxide resistors with resistance $\approx$~100~$k\Omega$ were added (for electrical isolation at high frequencies) using magnetron-sputtering, followed by gold wiring connections.  EBL patterning and lift-off were employed for both steps. A side-gate 3~$\mu$m from the nanowire was included for nanowire NbN100/6, whereas for nanowire NbN81/1, a parallel electrode 710~$\mu$m from the nanowire was employed as a side-gate. Further details of the fabrication process may be found in Refs.~\cite{Constantino2018} and~\cite{Fenton2016}. The designs and optical microscope images of the resulting devices are shown in Figs.~\ref{Fig3} and~\ref{Fig7}.

\section{Results}
\begin{figure}
\begin{center}
		\includegraphics[width=12cm]{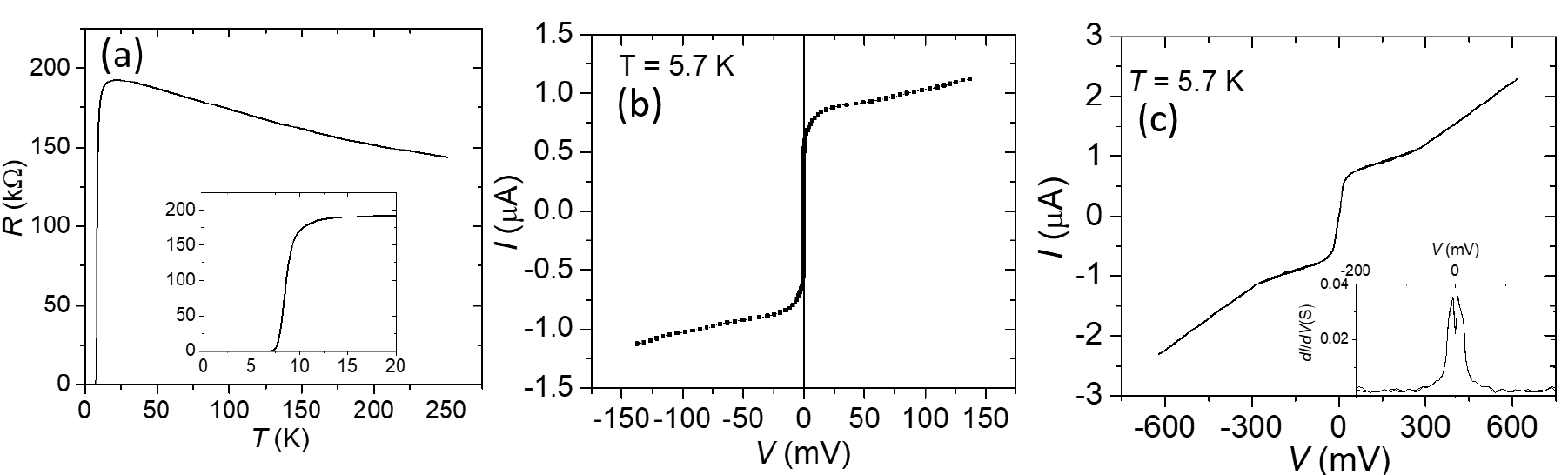}
		\caption{Characterisation of two nanowires on chip NbN81. (c) is a measurement on nanowire NbN81/1, as also shown in Figs.~\ref{Fig7},~\ref{Fig8} and~\ref{Fig9}. (a),(b) are measurements of a wider wire on the same chip, NbN81/2, showing normal superconducting behaviour for the film. (a) Variation of resistance with temperature for wire NbN81/2, using a source voltage of 100 mV; the inset shows the superconducting transition on an expanded scale. (b) $I$($V$) for wire NbN81/2. (c) $I$($V$) for nanowire NbN81/1. Low-bias departure from wide-wire superconducting behaviour is seen in $dI/dV$ (inset, obtained numerically from $I$($V$) close to the origin.}
		\label{Fig4}
	\end{center}
\end{figure}

Figure~\ref{Fig4} shows characterisation measurements on one of the two nanowire samples we have measured, NbN81. Measurements on both the nanowire NbN81/1 and a wider nanowire, NbN81/2, on the same chip show current-voltage characteristics with superconducting character at high bias, that is, for which a tangent to the slope at high bias ($\approx$~100~mV) intercepts the current axis at a positive value~\cite{Hongisto2012}. A superconducting transition with $T_{\rm c}$ around 9~K is found. These measurements verify that the film is superconducting. Further detail relating to characterisation of our nanowire samples may be found in Ref.~\cite{Constantino2018}.

\begin{figure}
\begin{center}
		\includegraphics[width=8cm]{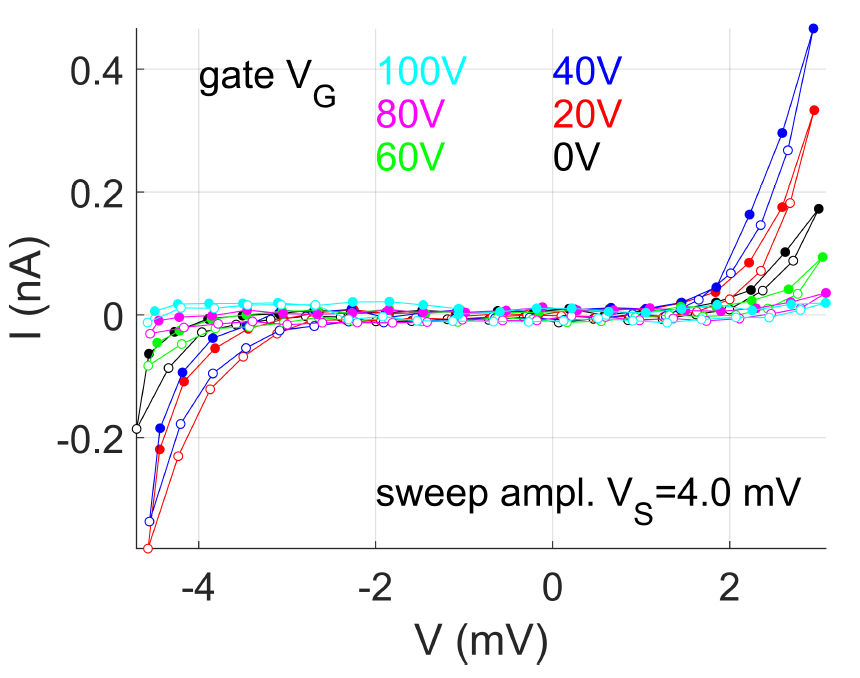}
		\caption{Nanowire current as a function of sample voltage for a range of fixed applied $V_{\rm g}$, at 300 mK, for nanowire NbN100/6. Filled (open) symbols indicate points taken while sweeping the sample voltage in the positive (negative) direction.}
		\label{Fig5}
	\end{center}
\end{figure} 

Current-voltage ($I$-$V$) measurements on nanowire NbN100/6 (on a second chip) were carried out at 300~mK and are shown in Fig.~\ref{Fig5}. With no applied $V_{\rm g}$, there is no current at low bias, with a $V_{\rm c}$ for the onset of conduction of 1.5~mV (positive sweep) and $-$ 3.5~mV (negative sweep), suggesting the presence of a voltage offset in the measurement with $V_{\rm c}$ around 2.5~mV when the offset is subtracted. When a voltage is applied to the top gate electrode, 3~$\mu$m from the nanowire (see Fig.~\ref{Fig3}(a) \& (b)) the characteristic $V_{\rm c}$ changes, decreasing as $V_{\rm g}$ increases to 20~V and then to 40~V, before increasing as Vg increases to 60~V, 80~V and 100~V.

\begin{figure}
\begin{center}
		\includegraphics[width=8cm]{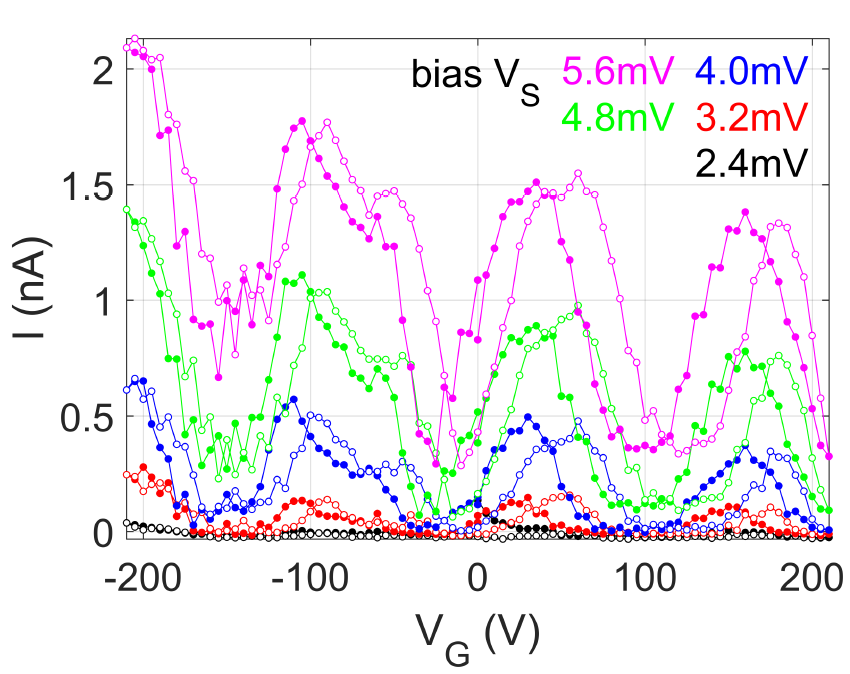}
		\caption{Nanowire current at constant voltage bias while sweeping $V_{\rm g}$, at 300~mK, for nanowire NbN100/6. Sweeps were conducted starting at 0~V, sweeping up to 220~V, then down to $-$220~V, then back to zero. Filled (open) symbols indicate points taken while sweeping the $V_{\rm g}$ in the positive (negative) direction. A small offset between the traces obtained as $V_{\rm g}$ is swept positively (left traces) compared with sweeping negatively (right traces) is an artefact, arising from very long $RC$ constants in the measurement system.}
		\label{Fig6}
	\end{center}
\end{figure} 

The periodic variation of the characteristics is more clearly seen when the dc voltage biasing the sample is fixed~\cite{quote3} and $V_{\rm g}$ swept. The current through the sample varies periodically as $V_{\rm g}$ is varied, as shown in Fig.~\ref{Fig6} for several different values of the dc voltage bias. Larger currents arise at $V_{\rm g}$s which have suppressed the $V_{\rm c}$, while smaller currents indicate a larger $V_c$.

A similar measurement was carried out on the other nanowire sample, NbN81/1. In this case, $V_{\rm g}$ was applied to a side-gate electrode much further from the nanowire, 710 $\mu$m away. The $IV$ characteristic shows similar behaviour: Fig.~\ref{Fig8} shows a dependence of the $IV$ on $V_{\rm g}$, with an apparent tuning of $V_{\rm c}$ due to $V_{\rm g}$. Fig.~\ref{Fig9} shows the variation of the current through the device at an applied voltage which biases the nanowire close to its $V_{\rm c}$. For this nanowire, there are again oscillations between maximum and minimum values, with a maximum of just over two full oscillations mapped out over the range of $V_{\rm g}$ applied, and the oscillation superimposed on a slower variation which may correspond to a second oscillation with much longer period~\cite{Hongisto2012} or may have another source. The particularly interesting feature in these measurements is apparent from a comparison of successive repetitions of the sweep of $V_{\rm g}$, all shown in Fig.~\ref{Fig9}. Successive sweeps were carried out starting at $V_{\rm g} = 0$ and increasing Vg in steps up to a maximum value before decreasing again to zero (sweeps 1$-$3, sweep 4 was terminated before $V_{\rm g}$ reached zero) or to a negative value of the same magnitude and then returning to zero (for sweeps 5$-$6). The successive sweeps were carried out to increasing values of the maximum $V_{\rm g}$ (indicated in the legend of Fig.~\ref{Fig9}). Each sweep took several minutes to collect. Despite all the measurements being essentially identical (other than the maximum value of $V_g$), three different dependences of the nanowire current on $V_{\rm g}$ are observed: at two points, between the first and second sweeps, and between the penultimate and last points of the fourth sweep, there is an abrupt change in the dependence being followed, which is then stable on the following sweep(s). The successive jumps each increase the rate of change of the nanowire current with $V_{\rm g}$. Complete periods of the oscillation are mapped out only with the most rapid Vg dependence (for sweeps 5$-$6 and the last point of sweep 4). However the strong similarity in curvature of this dependence to that of the preceding $V_{\rm g}$ dependence (for sweeps 2$-$4) suggests that this dependence represents a smaller part of the same variation, that is, that the switches between behaviour are simply changes in the period of the oscillations with $V_{\rm g}$. This is an important observation, because it gives insight into the physics responsible for the oscillations.

\section{Discussion}

\begin{figure}
\begin{center}
		\includegraphics[width=8cm]{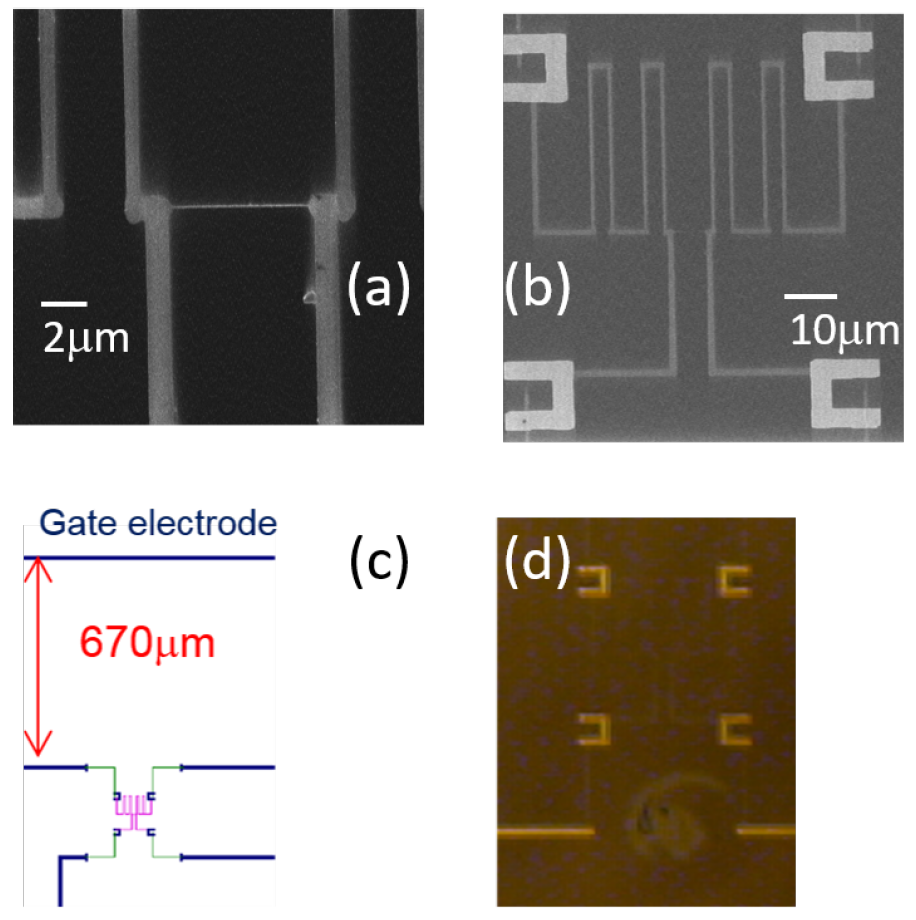}
		\caption{Nanowire NbN81/1. (a), (b) He-FIB micrographs of the central region of the device. (c) Design used to create the device. Pink elements are fabricated in NbN, while blue elements are gold. (d) Optical micrograph of the device. An electrically unconnected patch of NbN remaining after reactive-ion etching is visible below the device.}
		\label{Fig7}
	\end{center}
\end{figure} 

The behaviour observed for the two samples shows the oscillatory dependence on $V_{\rm g}$ expected for quantum interference between more than one CQPS element in series. For nanowire NbN100/6, in the oscillations of the current as the $V_{\rm g}$ is tuned [Fig.~\ref{Fig6}], no reduction in amplitude of oscillations is observed away from zero $V_{\rm g}$. This suggests that the comparison to the two-CQPS-element case is more appropriate than to a chain of more than two CQPS, in which a decay in the amplitude maxima would be expected as $V_{\rm g}$ increases. For a two-CQPS-element device, the period $\Delta V_{\rm g}$ of tuning of the device properties with an applied voltage is related by the gate$-$island capacitance $C_{{\rm g}i}$ and the change $\Delta q_i = 2e$ in the induced charge on the island by $\Delta q_i = 2e = C_{{\rm g}i} \Delta V_{\rm g}$. Applying this relationship to the period observed in the measurements shown in Fig.~\ref{Fig6}, we infer a capacitance $\approx$ 3 zF. However, for the geometry of NbN81/1, a simple estimation of the capacitance between the gate and one of the 1-$\mu$m-square islands is 4 aF, a discrepancy of three orders of magnitude. Notably, Hongisto and Zorin~\cite{Hongisto2012} reported a similar comparison between the capacitance inferred in this way and the value calculated from the geometry for two samples, finding good agreement in one case but, in the other, with the gate-voltage period being two orders of magnitude larger than expected based on the geometric capacitance (that is, the inferred capacitance is smaller than geometric value).

\begin{figure}
\begin{center}
		\includegraphics[width=8cm]{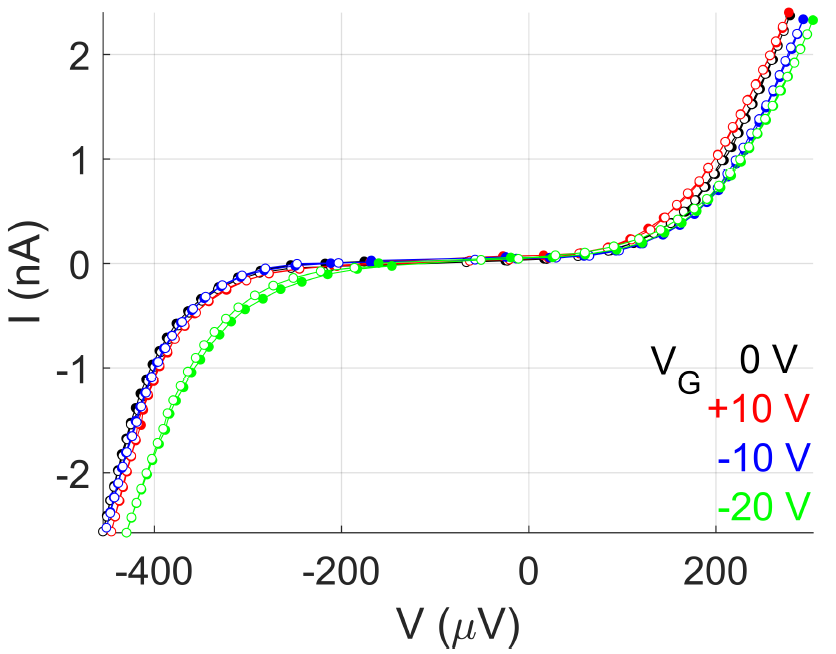}
		\caption{Nanowire current as a function of sample voltage for a range of fixed applied $V_g$, at 300 mK, for nanowire NbN81/1. Filled (open) symbols indicate points taken while sweeping the sample voltage in the positive (negative) direction.}
		\label{Fig8}
	\end{center}
\end{figure} 

\begin{figure}
\begin{center}
		\includegraphics[width=8cm]{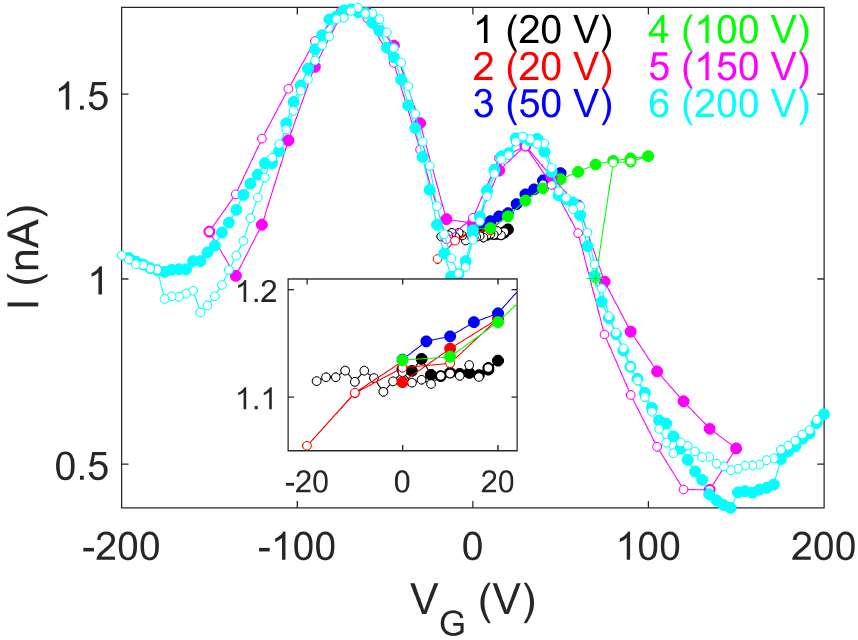}
		\caption{Nanowire current for nanowire NbN81/1, at 300~mK, at constant voltage bias while sweeping $V_{\rm g}$. Different colours indicate different sweeps at the same constant voltage bias, and are numbered in the legend in chronological order of collection, along with the maximum $V_g$ for the sweep. The inset shows the low $V_{\rm g}$ region on an expanded scale. Filled (open) symbols indicate the parts of the sweep where $V_{\rm g}$ was being increased (decreased).}
		\label{Fig9}
	\end{center}
\end{figure} 

The periodic behaviour observed in Fig.~\ref{Fig6} would be expected for CQPS interference between two identical elements with a single island. However, the geometry of sample NbN100/6 (Fig.~\ref{Fig3}) is actually four nanowire segments separated by three 1$-\mu$m$-$square islands, each with slightly different capacitance to the gate. The width of the nanowire also appears to be greater than the coherence length. These discrepancies may indicate that the properties of the nanowires arise from inhomogeneities in the sample rather than design features. If this is the case, the dimensions of the island could be 1 or 2 orders of magnitude smaller than 1~$\mu$m and therefore the capacitance to the gate correspondingly smaller; in other words, the discrepancy between the estimated and experimentally inferred capacitances is suggestive that inhomogeneity could be important, although inhomogeneity does not appear by itself to explain the whole of the discrepancy.

The oscillatory behaviour observed for nanowireNbN81/1 shows some differences to the NbN100/6 data: First, fewer complete oscillations are mapped out, secondly there is an apparent negative slope upon which the oscillations are superimposed, and thirdly the amplitude and period of the oscillations appear to change across the range studied. The clearly defined maximum and minimum features suggest that only one island or a few islands are involved, rather than many. Treating
the device as a two-element CQPS device, we infer an island-gate capacitance $\approx$~1 zF. By comparison, for an island of length 1~$\mu$m, the capacitance to the gate 710~$\mu$m away is expected to be $<$ 1 aF; this again shows a very large discrepancy to the inferred value. The apparently changing oscillation period indicates that the $V_{\rm g}$ may not be directly controlling the induced charge on the relevant island. Indeed, since there is no deliberately fabricated island and the length of the nanowire is more than two orders of magnitude greater than the $\xi_{\rm S}$, one expects either no islands (for a homogeneous nanowire) or many islands (many grain boundaries) rather than just one or a few as the well-defined variation between maximum and minimum implies.

The oscillatory nature of the behaviour can be explained to a similar level of agreement $-$ broad qualitative agreement, but not quantitative agreement $-$ by using a model based on a Josephson junction chain or network. Within a one-dimensional tunnel-junction chain model, the tuning of conduction in the chain should have a period $\Delta V = 2e/C_{{\rm g}i}$ (for Cooper pairs) in the case that all gate-island capacitances are the same, or otherwise be only quasi-periodic in other cases. We now turn to considering the abrupt jumps observed in the data in Fig.~\ref{Fig9}. Schollmann {\it et al}.~\cite{Schollmann2000} observed jumps between different values of CB/$V_{\rm c}$ and presented a model of a simple two-dimensional junction network which was capable of reproducing
the behaviour. Movements of charge in parts of the network shift the energy levels in other parts of the network which could reduce energy barriers which were preventing charge owing to reach the energetically most favourable state. While the details of the behaviour in Ref.~\cite{Schollmann2000} differ from our observed behaviour, it seems likely that the abrupt change in the tuning period in our observations may arise from a similar movement of some charges within the sample.

\section{Conclusion}
In conclusion, we have demonstrated modulation of electronic transport through superconducting nanowires by applying voltage to a nearby side-gate electrode. The behaviour shows features expected in the case of interference between coherent quantum phase slip elements, but also shows a similar level of consistency with an alternative picture based on CB effects in granular Josephson junctions. The wire length is more than two orders of magnitude greater than the superconducting coherence length, yet conduction periodic in $V_{\rm g}$ is observed. This suggests that there are correlations between the charges induced on the intervening islands. Furthermore, the absence of obvious decrease in amplitude of the current oscillations suggests quantum interference involves either one island or only a few islands. The observed switching behaviour suggests that control of the charge environment of the nanowire will be important for developing applications. The results highlight a live issue in the development of CQPS-based nanowire technologies, which is the need to categorically distinguish between these two different sources for the behaviour and call for further careful experimental work and possibly new theoretical ideas in order to resolve the issue.

\section*{Acknowledments}
This research was funded by the U.K. Engineering and Physical Sciences Research Council (EPSRC), grant number EP/J017329/1. The authors thank Oscar Kennedy for assistance with He-FIB imaging. We also acknowledge fruitful discussions with Louis Fry-Bouriaux.


\begin{thebibliography}{99}
\bibitem{quote1} The question of whether long nanowires are advantageous or undesirable for CQPS applications is also informed by suggestions that collective excitations in long wires might suppress CQPS.
\bibitem{quote2} The larger width is the result of a larger dose applied to this nanowire during the electron-beam lithography.
\bibitem{quote3} Note that, since the voltage bias drops across the series resistors, the sample voltage is lower than the applied bias voltage.
\end{thebibliography}
\end{document}